\documentclass{appolb}
\usepackage{epsfig}

\def\beq{\begin{equation}} 
\def\eeq{\end{equation}} 
\def\bea{\begin{eqnarray}} 
\def\eea{\end{eqnarray}} 

\def\non{\nonumber}

\def\MB{{\cal M}_B}
\def \ep{\epsilon} 
 
\def\mc{\mathcal} 
\def\lq{\left[} 
\def\rq{\right]}

\def\({\left(} 
\def\){\right)} 
\def\Re{\mathop{\rm Re}} 

\def\wwlndec{e^+\nu_e \mu^-\bar\nu_\mu}
\def\zzlldec{e^+e^- \,\mu^+\mu^-}
\def\zzlndec{e^+e^- \,\nu_\mu \bar\nu_\mu}

\begin{document}
\eqsec  
\title{QCD Corrections to Vector Boson Pair Production via Weak Boson Fusion
\thanks{Presented by B.\ J\"ager at {\it Physics at LHC}, Cracow, July 2006}%
}
\author{Barbara J\"ager and Dieter Zeppenfeld
\address{Institut f\"ur Theoretische Physik, 
        Universit\"at Karlsruhe, P.O.Box 6980, 76128~Karlsruhe, Germany}
\and
Carlo Oleari
\address{Dipartimento di Fisica "G. Occhialini", 
        Universit\`a di Milano-Bicocca, 
        20126~Milano, Italy}
}
\maketitle
\begin{abstract}
NLO-QCD corrections to vector boson pair production via weak boson fusion 
have recently been calculated and implemented into flexible parton-level 
Monte-Carlo programs. These allow for the computation of cross sections and
kinematical distributions within realistic experimental cuts. We summarize the
basic elements of the calculation and review phenomenological results 
for the LHC.
\end{abstract}
  
\section{Introduction}
One of the major goals of the CERN Large Hadron Collider (LHC) is the 
understanding of the mechanism of electroweak (EW) symmetry
breaking~\cite{ATLAS-CMS}. In this
context, vector-boson fusion (VBF) processes form a promising
class of reactions: Higgs production in VBF, i.e.\ the reaction $qq\to qqH$,  
has been proposed as a particularly clean channel for the 
discovery of the 
Higgs boson and a later determination of its couplings~\cite{Zeppenfeld:2000td}. 
An important background to the $H\to VV$ decay channel ($V=W$ or $Z$) 
in VBF is constituted by continuum $VV$ production via VBF, 
i.e.\ EW $pp\to VV jj$ production~\cite{wbfhtoww}. 
The precise knowledge of the standard model cross section for this reaction at
NLO-QCD accuracy becomes crucial for distinguishing enhancements in VBF processes
due to signatures of new physics, such as strong electroweak symmetry
breaking~\cite{sewsb},
from possible effects of higher order perturbative corrections.  

We have therefore calculated the NLO-QCD corrections to the reactions
$pp\to \wwlndec jj$~\cite{JOZ:WW}, $pp\to\zzlldec jj$, and 
$pp\to \zzlndec jj$~\cite{JOZ:ZZ}. In the following we will refer to these
processes briefly as ``EW $VVjj$ production''. Our calculations have been
turned into fully flexible parton-level Monte-Carlo programs allowing
for the computation of cross sections and distributions within realistic
experimental cuts, in complete analogy to the similar programs for the $Hjj$
signal~\cite{Figy:2003nv} and $Vjj$ 
production~\cite{Oleari:2003tc} via VBF. 
In this contribution we briefly recollect the elements of our calculation
and discuss the basic features of our results.

\section{Elements of the Calculation}
\label{sec:calculation}
\begin{figure}[t] 
\begin{center} 
\epsfig{figure=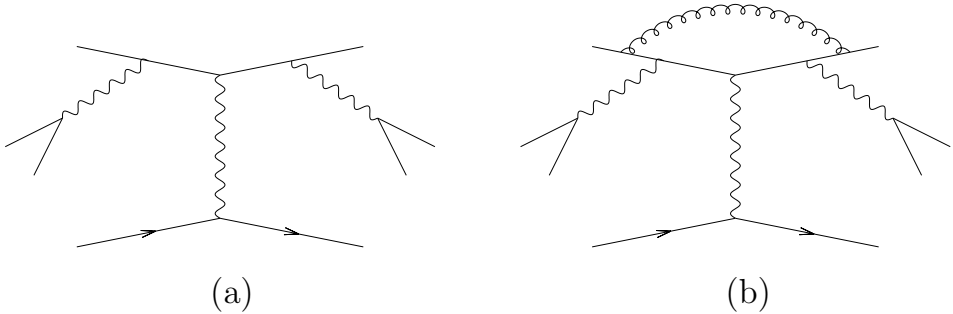,width=\textwidth,clip=} \ \  
\end{center} 
\caption
{\label{fig:feynBorn} 
Representative diagrams contributing to the LO cross section for the process
$qq\to qq \,\ell^+\ell^- \ell'^+\ell'^-$~(a) and to the virtual corrections 
for the same reaction~(b).
 }
\end{figure}
The computation of partonic matrix elements for EW $VVjj$ production is based on
the amplitude techniques of Ref.~\cite{HZ}. Special 
emphasis is put on the development of a numerically stable and fast code.  
That is achieved by organizing the calculation of the matrix elements in a modular way 
such that building blocks that are encountered in several diagrams are 
evaluated only once per phase space point, as described in some detail in 
Refs.~\cite{JOZ:WW,JOZ:ZZ}. 
This feature becomes particularly 
important for the computation of the real 
emission corrections to $VV jj$ production due to the large number of
contributing diagrams. 

Singularities emerging from soft and collinear configurations 
are regularized in the dimensional-reduction scheme~\cite{DR_citation} 
with space-time dimension $d=4-2\epsilon$. 
The cancellation of the divergences with the respective poles from the
virtual contributions is performed by introducing the 
counter-terms of the dipole subtraction method~\cite{CS}. 
The analytical form of the phase-space integrated 
subtraction terms, obtained after the factorization of the parton distribution
functions, is given by
\beq
\label{eq:I}
\langle \mc{I}(\ep)\rangle  = |\MB|^2 \frac{\alpha_s(\mu_R)}{2\pi} C_F
\left(\frac{4\pi\mu_R^2}{Q^2}\right)^\epsilon \Gamma(1+\epsilon)
\lq\frac{2}{\epsilon^2}+\frac{3}{\epsilon}+9-\frac{4}{3}\pi^2\rq\;,
\eeq
in the notation of Ref.~\cite{CS}, with $\MB$ denoting the amplitude of the
corresponding Born process and $Q^2$ 
the momentum transfer between the initial and final state quark in
Fig.~\ref{fig:feynBorn}. 
The computation of the virtual corrections requires the evaluation of
self-energy, triangle-, box-, and pentagon contributions on either the upper or
the lower quark line, as sketched in Fig.~\ref{fig:feynBorn}~(b) for one
representative diagram. 
Contributions from graphs  with gluons attached to both the
upper and lower quark lines vanish  at order $\alpha_s$.
Putting all virtual contributions together, we find 
\bea
\label{eq:virtual_born}
2 \Re \lq {\cal M}_V\MB^* \rq
&=& |\MB|^2 \frac{\alpha_s(\mu_R)}{2\pi} C_F
\(\frac{4\pi\mu_R^2}{Q^2}\)^\epsilon \Gamma(1+\epsilon)\\ \non
 &&\times
\lq-\frac{2}{\epsilon^2}-\frac{3}{\epsilon}+c_{\rm virt}\rq\
+2 \Re \lq \widetilde{\cal M}_V\MB^* \rq \,,
\eea
with $c_{\rm virt}=\pi^2/3-7$, and a completely finite remainder 
$\widetilde{\cal M}_V$. 
The divergent pieces in this expression exactly cancel the poles of the counter-terms 
in Eq.~(\ref{eq:I}). The remaining integrals are finite
and can be computed numerically in $d=4$ dimensions by means of a 
Passarino-Veltman tensor reduction~\cite{Passarino:1978jh}.  Numerical stability is achieved by the
repeated use of Ward identities which allow us to express a large fraction 
of the pentagon contributions by a combination of box-type diagrams. 

\section{Predictions for the LHC}
The cross-section contributions discussed above are implemented in fully
flexible parton level Monte-Carlo programs which allow us to calculate cross
sections and kinematical distributions for EW $WW jj$ and $ZZ jj$ production at
NLO-QCD accuracy within typical experimental acceptance cuts. 
We use the CTEQ6M parton distributions with
$\alpha_s(m_Z)=0.118$ at NLO, and the CTEQ6L1 set at LO~\cite{cteq6}. We chose
$m_Z=91.188$~GeV, $m_W=80.419$~GeV and 
$G_F=1.166\times 10^{-5}/$GeV$^2$ as electroweak input parameters, from which we
obtain $\alpha_{QED}=1/132.54$ and $\sin^2\theta_W=0.22217$. 
Jets are reconstructed from final-state partons employing the
$k_T$ algorithm~\cite{kToriginal,kTrunII} with resolution
parameter~$D=0.8$. Throughout, we set fermion masses to zero and neglect
external $b$- and $t$-quark contributions. 

In the following, we consider EW $ZZ jj$ production within generic cuts that are
relevant for VBF studies at the LHC. We require at least two hard jets with
\beq
\label{eq:cutspj}
p_{Tj} \geq 20~{\rm GeV} \, , \qquad\qquad |y_j| \leq 4.5 \, ,
\eeq 
where $y_j$ is the rapidity of the (massive) jet momentum which is reconstructed
as the four-vector sum of massless partons of pseudorapidity $|\eta|<5$. The two
reconstructed jets of highest transverse momentum are called ``tagging jets''.
To suppress backgrounds to VBF we impose a large rapidity separation of the two
tagging jets, 
\beq
\label{eq:cutsyjj}
\Delta y_{jj}=|y_{j_1}-y_{j_2}|>4\; .
\eeq
In addition, 
we require the two tagging jets to reside in opposite 
detector hemispheres,
\beq
\label{eq:cutsyj}
y_{j_1} \times y_{j_2} < 0\, ,
\eeq
with an invariant mass 
\beq
\label{eq:cutsmjj}
M_{jj} > 600~{\rm GeV}\;,
\eeq
and adopt the lepton cuts 
\bea
&& p_{T\ell} \geq 20~{\rm GeV} \,,\qquad |\eta_{\ell}| \leq 2.5  \,,\qquad 
\triangle R_{j\ell} \geq 0.4 \, , \nonumber \\
&& m_{\ell\ell} \geq 15~{\rm GeV} \,,\qquad  
\triangle R_{\ell\ell} \geq 0.2 \, ,
\label{eq:cutspl}
\eea
where  $\triangle R_{j\ell}$ and $\triangle R_{\ell\ell}$ denote the 
jet-lepton and lepton-lepton separation in the rapidity-azimuthal angle plane,
respectively, 
and $m_ {\ell\ell}$ the invariant mass of an $e^+ e^-$ or $\mu^+ \mu^-$ pair. 

The total cross section for EW $VVjj$ production contains contributions from the
Higgs resonance with $H\to VV$ decays, as well as from $VV$ continuum production. 
In the following we will focus on the $VV$ continuum by 
imposing a cut on the four-lepton invariant mass
\beq
\label{eq:offres}
M_{VV} = \sqrt{(p_{\ell_1}+p_{\ell_2}+p_{\ell_3}+p_{\ell_4})^2} > 
m_H+10\;{\rm GeV}\;,
\eeq
where the $p_{\ell_i}$ denote the four-momenta of the leptons 
produced in the specific reaction under consideration.

\begin{figure}[!tb] 
\centerline{ 
\epsfig{figure=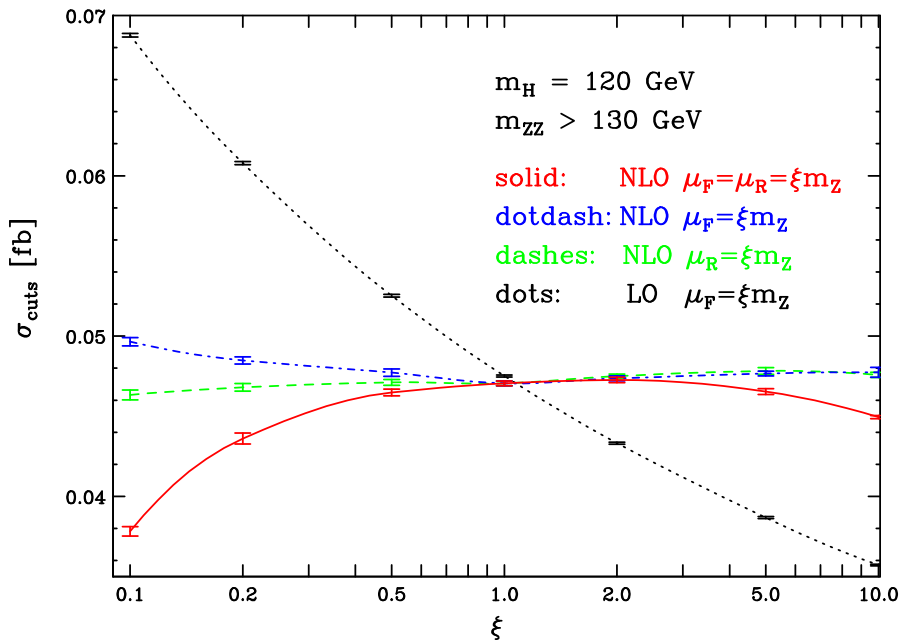,width=0.95\textwidth,clip=}
} 
\caption
{\label{fig:scale-dep} 
Dependence of the total  $pp\to \zzlldec jj$  continuum cross section at
the LHC on the factorization and renormalization scales.
}
\end{figure}

The total continuum cross section for the reaction $pp\to \zzlldec jj$  
at NLO within the cuts of Eqs.~(\ref{eq:cutspj}-\ref{eq:offres}) and a Higgs 
mass of $m_H=120$~GeV is displayed in Fig.~\ref{fig:scale-dep}
as a function of the renormalization scale $\mu_R=\xi m_Z$ with a fixed 
factorization scale $\mu_F=m_Z$ (dashed green curve), as a function of 
$\mu_F=\xi m_Z$ with $\mu_R=m_Z$ (dot-dashed blue curve), and for the case, 
where both scales are varied simultaneously, 
$\mu_R=\mu_F=\xi m_Z$ (solid red curve). The LO cross section depends only on
$\mu_F=\xi m_Z$ (dotted black curve).  
As in the analogous cases of $\wwlndec$ and
$\zzlndec$ production via VBF, discussed in Refs.~\cite{JOZ:WW,JOZ:ZZ}, the
scale variations of the NLO prediction are found to be 
below the 2\% level when the scale parameter runs from $\xi = 0.5$ to 
$\xi = 2.0$, while the LO cross section changes by more than 20\% in the 
same range of $\xi$.

\begin{figure}[!tb] 
\centerline{ 
\epsfig{figure=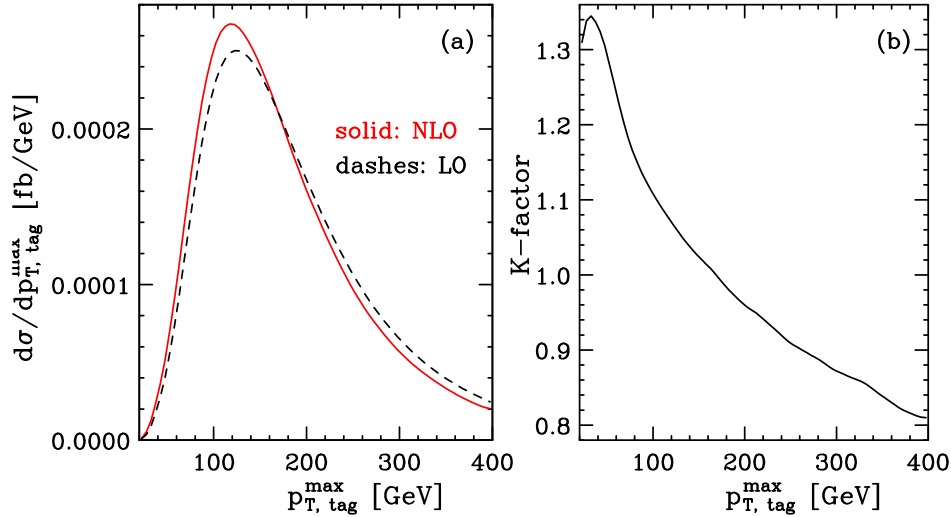,width=\textwidth,clip=}
} 
\caption
{\label{fig:ptmax_tag} 
Transverse momentum distribution of the tagging jet with the highest $p_T$ in EW
$\zzlldec jj$ production at the LHC. Panel~(a) shows the NLO result (solid red
line) and the LO prediction (dashed black line). Panel~(b) displays the
$K$~factor defined in Eq.~(\ref{eq:kfac}).
}
\end{figure}

\begin{figure}[!tb] 
\centerline{ 
\epsfig{figure=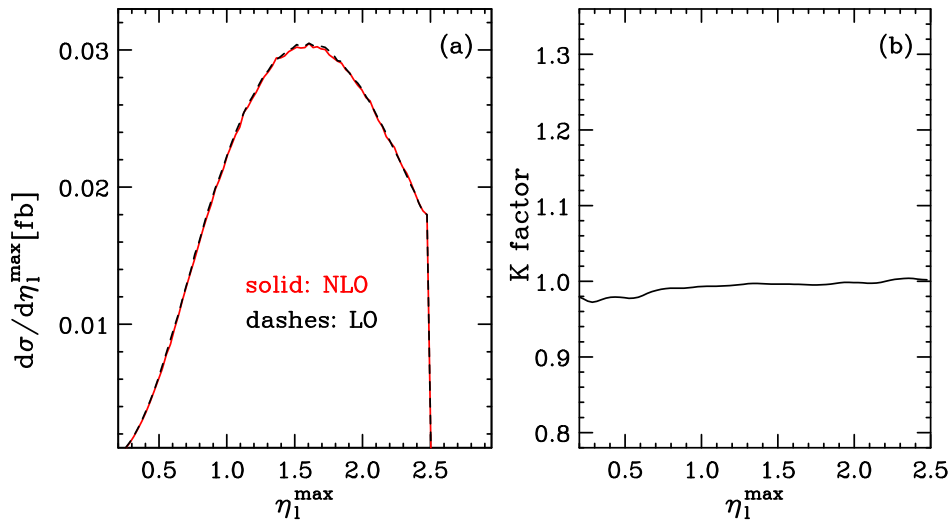,width=\textwidth,clip=}
} 
\caption 
{\label{fig:ylmax} 
Distribution of the maximal lepton rapidity in  EW $\zzlldec jj$ production
at the LHC. Curves are as is Fig.~\ref{fig:ptmax_tag}.
}
\end{figure}

Figure~\ref{fig:ptmax_tag}
shows the transverse momentum distribution of the tagging jet with the highest
$p_T$ for EW $\zzlldec jj$ production. The
factorization and renormalization scales are fixed to $\mu_F=\mu_R=m_Z$. The
$K$~factor, defined by
\beq
\label{eq:kfac}
K(x) = \frac{d\sigma_{NLO}/dx}{d\sigma_{LO}/dx}
\eeq
and shown for this distribution in Fig.~\ref{fig:ptmax_tag}~(b), illustrates 
the change in shape when going from LO to NLO. At NLO, smaller values of $p_T$
are preferred by the tagging jet, which is due to the extra parton which emerges
in the real emission contributions. 
In a similar way, other distributions such as the invariant mass distribution of
the tagging-jet pair, $d\sigma/dM_{jj}^{tag}$~(see Ref.~\cite{JOZ:ZZ}), or the
transverse momentum distributions of the detected leptons are affected by the
inclusion of NLO corrections. 
On the other hand, some angular distributions such as $d\sigma/d\eta_\ell^{max}$, where
$\eta_\ell^{max}$ designates the largest lepton rapidity emerging in the
scattering,
barely change at NLO as illustrated in Fig.~\ref{fig:ylmax}. 
The step in $d\sigma/d\eta_\ell^{max}$ at $\eta_\ell^{max}=2.5$ is
due to the rapidity cut imposed on the lepton rapidities in~Eq.(\ref{eq:cutspl}).

Our discussion on the effect which higher order QCD corrections 
have on dynamical $K$~factors has been based on EW $\zzlldec jj$ production.  
However, the shapes of kinematical distributions 
in the other VBF production processes we have studied, i.e., 
$pp\to \zzlndec jj$ and $pp\to \wwlndec jj$, behave very similarly.

\section{Conclusions}
In this contribution we have reviewed some results on EW $WWjj$ and $ZZjj$ 
production at NLO-QCD accuracy obtained with fully flexible Monte-Carlo 
programs. 
The higher order corrections to these reactions turn out to be under excellent
control, as indicated by the small scale dependence of the total cross sections
and $K$~factors close to one. We found, however, that some kinematical
distributions exhibit a noticeable change in shape when NLO corrections are
considered. This indicates the importance of including NLO-QCD contributions in
precision studies of VBF processes at the LHC. 

\section*{Acknowledgements}
This research was supported in part by the Deutsche Forschungsgemeinschaft
under SFB TR-9 ``Computergest\"utzte Theoretische
Teilchenphysik''. 


\begin{thebibliography}{99}
\bibitem{ATLAS-CMS}
ATLAS Collaboration, ATLAS TDR,
Report No.\ CERN/LHCC/99-15 (1999);
G.~L.~Bayatian {\it et al.}, CMS Technical Proposal,
Report No.\ CERN/LHCC/94-38x (1994).

\bibitem{Zeppenfeld:2000td}
D.~Zeppenfeld, R.~Kinnunen, A.~Nikitenko and E.~Richter-Was,
Phys.\ Rev.\ D {\bf 62}, 013009 (2000)
[arXiv:hep-ph/0002036];
D.~Zeppenfeld,
in {\it Proc. of the APS/DPF/DPB Summer Study on the Future of
Particle Physics, Snowmass, 2001} edited by N.~Graf,
eConf {\bf C010630}, p.\ 123 (2001)
[arXiv:hep-ph/0203123];
A.~Belyaev and L.~Reina,
JHEP {\bf 0208}, 041 (2002)
[arXiv:hep-ph/0205270];
M.~D\"uhrssen et al.,
Phys.\ Rev.\ D {\bf 70}, 113009 (2004)
[arXiv:hep-ph/0406323].

\bibitem{wbfhtoww}
D.~Rainwater and D.~Zeppenfeld,
Phys.\ Rev.\ D {\bf 60}, 113004 (1999)
[Erratum-ibid.\ D {\bf 61}, 099901 (2000)]
[arXiv:hep-ph/9906218];
N.~Kauer, T.~Plehn, D.~Rainwater and D.~Zeppenfeld,
Phys.\ Lett.\ B {\bf 503}, 113 (2001)
[arXiv:hep-ph/0012351].


\bibitem{sewsb} 
J.~Bagger {\it et al.},
Phys.\ Rev.\ D {\bf 52} 3878 (1995)
[arXiv:hep-ph/9504426];
M.~S.~Chanowitz,
Czech.\ J.\ Phys.\  {\bf 55}, B45 (2005)
[arXiv:hep-ph/0412203].

\bibitem{JOZ:WW}
B.~J\"ager, C.~Oleari and D.~Zeppenfeld,
JHEP {\bf 0607}, 015 (2006) 
[arXiv:hep-ph/0603177].

\bibitem{JOZ:ZZ}
B.~J\"ager, C.~Oleari and D.~Zeppenfeld,
Phys.\ Rev.\ D {\bf 73} 113006 (2006)
[arXiv:hep-ph/0604200].

\bibitem{Figy:2003nv}
T.~Figy, C.~Oleari and D.~Zeppenfeld,
Phys.\ Rev.\ D {\bf 68}, 073005 (2003)
[arXiv:hep-ph/0306109].

\bibitem{Oleari:2003tc}
C.~Oleari and D.~Zeppenfeld,
Phys.\ Rev.\ D {\bf 69}, 093004 (2004)
[arXiv:hep-ph/0310156].


\bibitem{HZ}
K.~Hagiwara and D.~Zeppenfeld,
Nucl.\ Phys.\ {\bf B274}, 1 (1986);
K.~Hagiwara and D.~Zeppenfeld,
Nucl.\ Phys.\  {\bf B313}, 560 (1989).

\bibitem{DR_citation}
Warren Siegel, Phys.\ Lett.\ B {\bf 84}, 193 (1979);
Warren Siegel, Phys.\ Lett.\ B {\bf 94}, 37 (1980).

\bibitem{CS}
S.~Catani and M.~H.~Seymour,
Nucl.\ Phys.\  {\bf B485}, 291 (1997)
[Erratum-ibid.\  {\bf B510}, 503 (1997)]
[arXiv:hep-ph/9605323].

\bibitem{Passarino:1978jh}
G.~Passarino and M.~J.~Veltman,
Nucl.\ Phys.\ {\bf B160}, 151 (1979).


\bibitem{cteq6}
J.~Pumplin, D.~R.~Stump, J.~Huston, H.~L.~Lai, P.~Nadolsky and W.~K.~Tung,
JHEP {\bf 0207}, 012 (2002)
[arXiv:hep-ph/0201195].

\bibitem{kToriginal}
S.~Catani, Yu.~L.~Dokshitzer and B.~R.~Webber,
                Phys.\ Lett.\ B  {\bf 285} 291 (1992);
S.~Catani, Yu.~L. Dokshitzer, M.~H.~Seymour and B.~R.~Webber,
                Nucl.\ Phys.\  {\bf B406} 187 (1993);
S.~D.~Ellis and D.~E.~Soper, Phys.\ Rev.\ D
{\bf 48} 3160 (1993).

\bibitem{kTrunII}
G.~C.~Blazey {\it et al.},
arXiv:hep-ex/0005012.

\end{thebibliography}
\end{document}